# The First Cobalt Single-Molecule Magnet

En-Che Yang and David N Hendrickson

Department of Chemistry and Biochemistry, University of California at San Diego, La Jolla, CA 92037, USA

Wolfgang Wernsdorfer

L. Neel-CNRS BP 166, 25 Avenue des Martyrs 38042 GRENOBLE Cedex 9, France

Motohiro Nakano

Department of Molecular Chemistry, Graduate School of Engineering, Osaka University, 1-16 Machikaneyama, Toyonaka, Osaka 560-0043, Japan

Roger Sommer and Arnold L Rheingold

Department of Chemistry and Biochemistry, University of Delaware, Newark, Delaware 19716

Marisol Ledezma-Gairaud and George Christou

Department of Chemistry, University of Florida, Gainesville, Florida 32611-7200

The first cobalt molecule to function as a single-molecule magnet, [$Co_4(hmp)_4(MeOH)_4Cl_4$], where hmp$^-$ is the anion of hydroxymethylpyridine, is reported. The core of the molecule consists of four Co(II) cations and four hmp$^-$ oxygen atoms ions at the corners of a cube.  Variable-field and variable-temperature magnetization data have been analyzed to establish that the molecule has a S=6 ground state with considerable negative magnetoanisotropy. Single-ion zero-field interactions ($DS_z^2$) at each cobalt ion are the origin of the negative magnetoanisotropy. A single-crystal of the compound was studied by means of a micro-SQUID magnetometer in the range of 0.040-1.0K. Hysteresis was found in the magnetization versus magnetic field response of this single crystal. It is concluded that this is the first cobalt molecule to function as a single-molecule magnet.

01.30.Cc

Session No. CP-01

**Introduction**

Single-molecule magnets (SMM's) are of great interest because they provide a means to systematically study the chemistry and physics of nanomagnets[1]. They can be considered as slowly relaxing magnetic particles of well-defined size, shape and orientation. A single-molecule magnet has a large-spin ground state with appreciable negative magnetic anisotropy (Ising type), resulting in an energy barrier ($DS_z^2$) for the reversal of the direction of magnetization. At low temperatures, magnetization hysteresis loops and out-of-phase ac susceptibility signals are seen.

Even though many SMM's have been synthesized, only one molecule has been reported for a transition metal ion with more than five d electrons[2]. If a molecule is to be a SMM, it must have a negative magnetoanisotropy (D<0 for the $DS_z^2$ interaction) and metal ions with $d^n$(n>5) typically have D>0. It is known that for molecules with two or more metal ions that the overall magnetoanisotropy is the vectorial projection of the single ion contributions[3]. Theoretically, it is possible for a molecule to have D<0 even though the constituent metal ions have single-ion zero-field interactions with D>0. In this paper we report the first cobalt SMM. It is known that Co(II) ions have D>0.

**Experimental Results**

The new complex with composition [Co$_4$(hmp)$_4$(MeOH)$_4$Cl$_4$] (called Co$_4$), where hmp$^-$ is the anion of hydroxymethylpyridine, was prepared by reacting Co(H$_2$O)$_6$Cl$_2$, sodium methoxide (MeONa), and hydroxymethylpyridine (1:1:1 ratio) in methanol. The single crystal X-ray structure of this complex has been determined. The Co$_4$ molecule crystallizes in tetragonal space group $I\bar{4}2b$. The core of this molecule consists of four Co(II) and four hmp$^-$ at the corners of a cube [4]. The detailed structure data will be discussed in latter report. The fact that this complex has S4 site symmetry for an individual molecule in the crystal suggests that the orthogonal hard-axis alignment[5] for the cubane type metal cluster will be suitable for us to analyze this case.

The magnetic susceptibility ($\chi_M$) of Co$_4$ has been measured in a 10kG dc field and in a 1.0 G ac field oscillating at 1000Hz. The $\chi_M T$ value is 11.8 cm$^3$ mol$^{-1}$ K at 300 K and remains essentially constant as the temperature is decreased until ca. 30 K when it begins to increase to a maximum value of 12.4 cm$^3$ mol$^{-1}$ K at 15 K before rapidly decreasing to 5.4 cm$^3$ mol$^{-1}$ K at 2 K. The out-of-phase ac susceptibility ($\chi_M''$) is less than 0.01 cm$^3$ mol$^{-1}$ at temperatures above 3.5 K and substantially increases to 0.16 cm$^3$ mol$^{-1}$ as the temperature is decreased to 1.8 K. The substantial increase of the out-of-phase ac signal suggests that the Co$_4$ molecule has an appreciable energy barrier for reversing its magnetization.

The Co$_4$ complex was determined to have a large ground state spin by an analysis of the two types of low-temperature, variable-field magnetization data depicted in Figures 1 and 2. Figure 1 gives data for a sample aligned with the magnetic field, whereas Figure 2 illustrates the random powder sample data. The simulation of both types of data sets was carried by employing the spin Hamiltonian for an orthogonal hard-axis type alignment model[5] as given in eq. 1:

$$\hat{H} = -2J_0(\hat{S}_2\hat{S}_3 + \hat{S}_4\hat{S}_1 + \hat{S}_1\hat{S}_3 + \hat{S}_2\hat{S}_4) - 2J_1(\hat{S}_1\hat{S}_2 + \hat{S}_3\hat{S}_4) + D'(\hat{S}_{1x}^2 + \hat{S}_{2y}^2 + \hat{S}_{3x}^2 + \hat{S}_{4y}^2) + g\mu_B B\hat{S}_z \quad (1)$$

In this equation $S_n$ (n=1,2,3,4) is the spin on each metal atom n, $S_{nj}$ (j=x,y,z) is the component of $S_n$ in the j direction, D' is the axial zero-field splitting for each individual metal atom, and $J_0$ and $J_1$ are the exchange coupling constants. The magnetization of a polycrystalline sample of Co$_4$ is given by the Van Vleck eq 2 :

$$M = N \times [\sum(-\partial E_i / \partial H)\exp(-E_i/kT) / \sum \exp(-E_i/kT)] \quad (2)$$

In the above equation N is Avogadro's number, k is the Boltzmann constant and $\partial E_i/\partial H$ is the change in energy of the *i*th level in response to a change in the magnetic field. Full matrix diagonalization of the Hamiltonian matrix was carried out in the uncoupled basis set. The zero-field splitting in the ground state of the Co$_4$ molecule (D) is related to the single-ion zero-field parameter (D') as given in eq 3.

$$D = -(1/8)D' \quad (3)$$

The best simulation was found with g= 1.97, D'= 44.2 K, $J_0$= 5.15 K, $J_1$= -6.95 K for Figure 1, and g= 2.12, D'= 44.8 K, $J_0$= 4.77 K, $J_1$= -7.80 K for the data in Figure 2. These two sets of parameters lead to the result that $S_T$=6 ($S_T$ is the total spin for the whole molecule) and the resultant axial zero-field splitting for the whole molecule is D ≅ -5.6 K.

We used the micro-SQUID technique[6] to measure magnetization hysteresis loops of a single crystal for complex **1** at different temperatures and field sweep rates. The field was roughly aligned with the easy axis of magnetization. Figure 3 shows hysteresis loops with a constant field sweep rate of 0.14 Tesla/sec at different temperatures ranging from 0.04 K to 1.1 K. The loops present hysteresis for temperatures below about 1.2 K which increase rapidly upon decreasing the temperature. This is expected for a SMM. On the other hand, Figure 4 shows hysteresis loops at a constant temperature of 0.04K for different field scanning rates between 0.002 Tesla/sec and 0.140 Tesla/sec. The hysteresis loops only weakly depend on the field scanning rate which suggests that resonant quantum tunneling is hindered by small intermolecular interactions. More studies are in progress to improve the alignment of the fields and to understand better the fine structure of the hysteresis loops.

## Conclusion

The $Co_4$ molecule has been found to be a single-molecule magnet. This is the first cobalt SMM and establishes that this phenomenon can occur with molecules comprised of metal atoms with single-ion zero-field interactions ($DS_z^2$) where D>0.

## Acknowledgments

D.N.H. and G.C. thank the National Science Foundation for support of this research.

**Figure1. Plot of reduced magnetization measurement with sample aligned with magnetic field and the simulation curve.**

**Figure2. Plot of reduced magnetization measurement of randomly oriented powder sample and the simulation curve.**

**Figure3. Plot of hysteresis loop with constant scanning rate (0.140 T/sec) at four different temperatures**

**Figure4. Plot of hysteresis loop with different scanning rates at a constant temperature (0.04K).**

[1] L. Gunther, Phys. World , December 28 (1990).

[2] H. Oshio, N. Hoshino, and T. Ito, J. Am. Chem. Soc **122**, 12602-12603 (2000).

[3] D. Gatteschi and L. Sorace, J. Solid State Chem **159(2)**, 253-261 (2001).

[4] A. Escuer, M. Font-Bardýa , S. B. Kumar, X. Solans, and R. Vicente, Polyhedron **18**, 909-914 (1999).

[5] M. Nakano, G. Matsubayashi, T. Muramatsu, T. C. Kobayashi, K Amaya, J. Yoo, G. Christou, and D. N. Hendrickson, Mol. Cryst. Liq. Cryst , in press (2001).

[6] W. Wernsdorfer, Adv. Chem. Phys. **118**, 99 (2001).

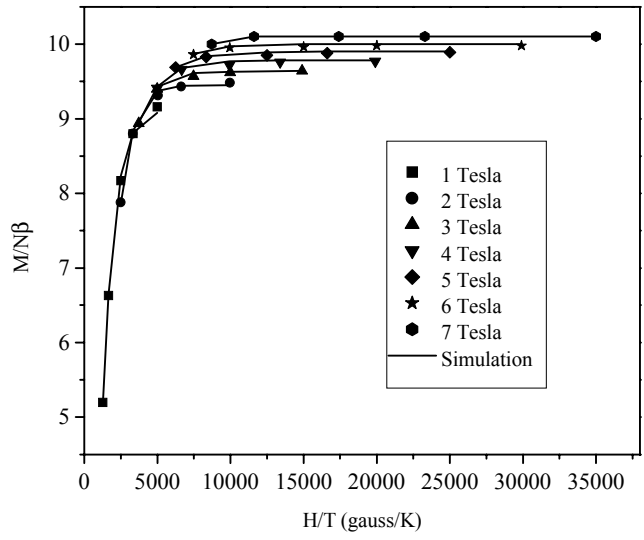

Figure 1
En-Che Yang
Journal of Applied Physics

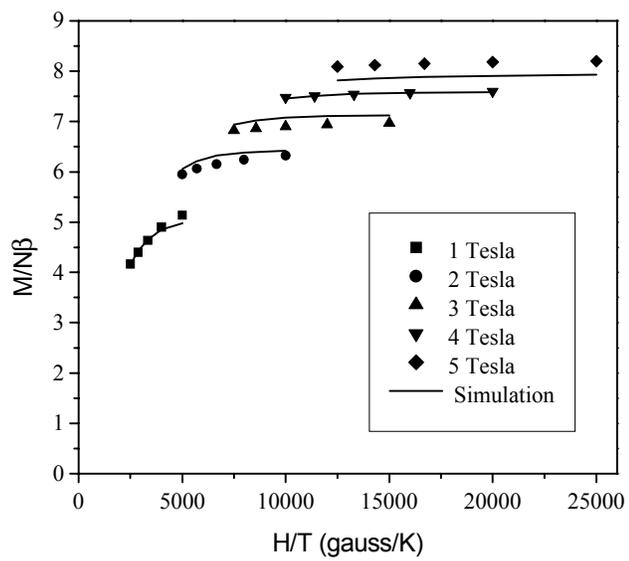

Figure 2
En-Che Yang
Journal of Applied Physics

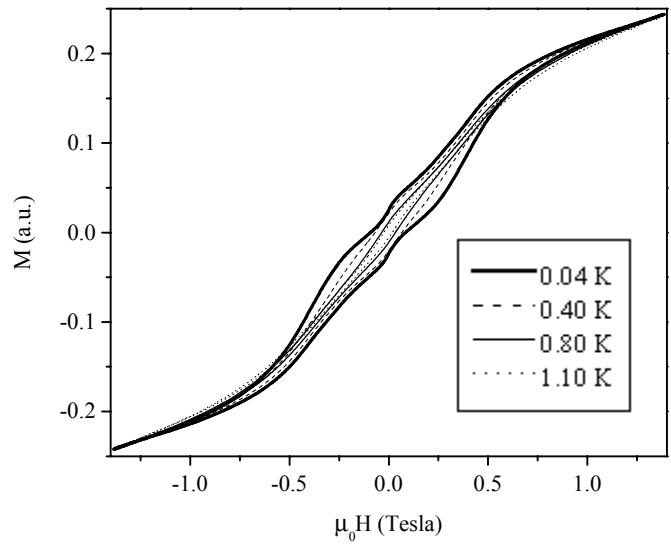

Figure 3
En-Che Yang
Journal of Applied Physics

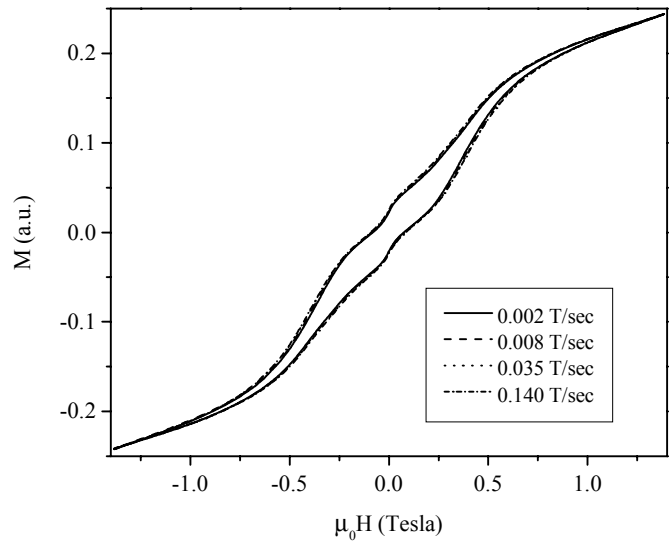

Figure 4
En-Che Yang
Journal of Applied Physics